\documentclass[preprint,double-spaced,aps,apssymb]{revtex4}
\usepackage{amsmath,amssymb}
\usepackage{graphicx}
\usepackage{dcolumn}
\usepackage{bm}
\usepackage{psfrag}
\newcommand{\ber}{\begin{eqnarray}}
\newcommand{\eer}{\end{eqnarray}}
\newcommand{\bea}{\begin{equation}}
\newcommand{\eea}{\end{equation}}

\begin{document}

\title{Model of amplitude modulations induced by phase slips in one-dimensional superconductors}

\author{A. Bhattacharyay}
\email{a.bhattacharyay@iiserpune.ac.in}
 \affiliation{Indian Institute of Science Education and Research, Pune, India}

\date{\today}

\begin{abstract}
We propose a linear model for the dynamics of amplitudes associated with formation of phase slip centers in a one-dimensional superconductor. The model is derived taking into account the fact that, during the formation of phase slip centers the wave number of the superconducting phase remains practically constant. The model captures various forms of amplitude modulations associated with PSCs in closed analytic forms.
\end{abstract}
\maketitle
One dimensional (1D) superconductor has its width smaller than the Ginzburg-Landau (G-L) coherence length $\xi(T)$ (average separation between electrons in a cooper pair) where T is the temperature of the system. Near the critical temperature $T_C$ of the superconducting (SC) to normal (N) phase transition, the coherence length $\xi(T)$ is quite large and a finite width channel or wire of superconducting material can be treated as a 1D system. Such a system has the pure SC state (characterized by having no Ohmic voltage drop along the wire despite having current through it) globally stable below a critical current density $j_c$ (G-L critical current) at $T<T_C$. At a current density $j>j_c$, being induced by an applied voltage across the length of the wire, the SC state coexists with the N state up to an upper limiting current $j_2$. Beyond $j_2$ the N state is globally stable. Within the limits $j_c<j<j_2$, the SC state is metastable and gives up locally to the N state by occasional formation of phase slip centers (PSCs) \cite{lan,mac,kop}. PSCs are the points in space where the amplitude ($A$) of the SC order parameter becomes zero. By the formation of PSCs, the complex order parameter of the SC state loses (or adds) a turn along the length of the wire and moves on to a state characterized by a smaller (or larger) wave number ($q$) to generally lower the free energy ($F$) \cite{lan,mac,kop,grp1}. Many localized forms of the SC amplitude and corresponding electrochemical potential $\mu$ (N state order parameter in the present context) have mostly been numerically observed in this regime \cite{kra,kop,pee1,pee2,pee3} under various circumstances. For 1D superconductors, although it is an old classic problem, new exact results which were unknown previously are constantly being explored \cite{zha} in view of the vast applicational importance of the subject. Despite so much work being done, till date, there is no single relationship that can at least partially capture the function space of the localized amplitude variations and corresponding chemical potential profiles in a general way for the resistive state. The present work is focused at achieving this goal.
\par
In this letter we propose a model to study the amplitude modulations associated with PSC formation. We extend the notion of PSC formation, as envisaged by Little \cite{lit}, to hypothesize that, the wave number of the SC phase during a PSC formation remain constant till the amplitude vanishes locally. This is the main physical observation which would immediately imply, the corresponding phenomenological G-L free energy is effectively function of a single variable (amplitude of the SC phase). The simple model, we propose here, will be based on this new observation of effective uni-dimensional G-L free energy. We will solve this model to explicitly get the closed analytical form for dynamic and static PSCs and other amplitude modulations. The above mentioned fundamental hypothesis of the constancy of the wave number during local variations of the amplitude allows us to derive a range of amplitude modulated SC and corresponding N phase profiles on the same footing of a central relationship that relates these quantities. Such a unified derivation of a range of exact analytic results are remarkable in view of the prevailing notion of the analytic intractability of the full nonlinear model (CGL) for such solutions. This point gets justified when we provide analytic expressions for numerically observed amplitude modulations by Kramer et al \cite{kra} in 1977 which, to our knowledge, has not analytically been accounted for till date. 
\par
The dynamics of a 1D superconductor in the resistive regime is given by CGL equation \cite{mac} as 
\ber\nonumber
\psi_t+i\mu\psi &=& \psi_{xx} + (1-\vert\psi\vert^2)\psi \\
j&=&Im(\psi^*\bigtriangledown\psi)-\mu_x .
\eer    
Here, we consider, the simplest possible form of the time dependent CGL to demonstrate our novel method of exploring the function space of localized amplitude modulations. In what follows, the treatment would be general, disregarding details of the dependencies on the coefficients and that is why we have set most of the coefficients to unity in Eq.1. In this model, $\psi$ is the superconducting order parameter which is complex valued. The system being 1D, $x$ is distance along the wire from some arbitrary origin. Let us consider, the length of the wire is L with cross sectional area $\sigma$. The $\mu$ is electrochemical potential which can be considered as the order parameter of the N phase within the scope of G-L phenomenology. The $j$ is current (density) through the superconducting wire and the suffix $t$ and $x$ of $\psi$ indicate of the partial derivatives of the SC order parameter with respect to them. 
\par
Eq.1 has two stationary solutions. (1) $\psi \equiv 0 $, $\mu = -xj$ (for constant $j$) which is the normal state and (2) $\psi=Ae^{iqx}$, $q^2=1-A^2$, $j=A^2q$, which is the superconducting state when $\mu\equiv 0$ \cite{kra}. The SC order parameter can be visualized as a spiral wound around the x-axis (along the wire). If there are N turns along the length L, there is a total phase difference $\phi = 2\pi N$ along L. Thus, the wave number q of the SC phase is a measure of number of turns the system has on a given length L, since, $q=2\pi N/L$. The expression for the corresponding G-L free energy, when the steady state solution is purely superconducting, is given by 
\bea
F= L\sigma[(q^2-1)A^2+\frac{A^4}{2}].
\eea     
The form of F clearly indicates, the SC states with smaller q values are energetically favored. In other words, the spiraling SC order parameter would tend to lose its turns along the length of the sample to go to a lower free energy state. The turn loss happens when the amplitude of the SC phase locally vanishes and this local vanishing of the amplitude associated with a phase change of $2\pi$ is what known as PSC formation. The qualitative idea of PSC formation was originally given by Littel \cite{lit}. Little had pointed out that the boundary conditions make $\psi$ single valued and any internal (not at boundary) change of phase has to be achieved by a part of the spiral crossing the central axis that means a local vanishing of the amplitude. 
\par
We argue that, so long as the amplitude does not exactly become zero at a point, while forming the PSC, the wave number of the SC phase remains necessarily unaltered because number of turns along the length of the sample remains unaltered. The width of a spiral can always shrink locally keeping turns unaltered. There is no apparent geometric reason as to why it should not happen. Moreover, the above mentioned amplitude wave number relationship is a characteristic feature of only the stationary SC state and it does not ensure a coupling between the amplitude and the wave number while a PSC forms. Our argument necessarily means a separation of the dynamics of the wave number and amplitude of the SC phase. So, we extend Little's idea to put forward the hypothesis that, the phase relaxation only happens after the phase change is in place somewhere locally along the wire and amplitude and phase dynamics are independent of each other during such a process. 
\par
At this stage, refer to Fig.1 where a schematic diagram of the free energy F has been plotted against amplitude of the SC phase for the states with wave numbers $q_0$ (upper branch) and $q_0-2\pi/\lambda$ (lower branch) where $\lambda$ is the wavelength of the SC phase with 1 less turn than that having a wave number $q_0$. During this process of formation of PSC, the system will move uphill (induced by thermal fluctuations or whatever) from P to Q (along the upper branch in Fig.1) where the amplitude actually becomes zero. Now, being at Q the SC phase can only lose one turn and can eventually take the lower branch to come down to the minimum at R corresponding to the wave number $q_0-2\pi/\lambda$. Interesting to note that, the assumption of constancy of the wave number essentially makes the free energy defined in one dimensional space and thus, there are no other paths available for the system to move to the lower branch while forming PSCs. 
\par
Therefore, through out this transition, the wave number remains fixed either at $q_0$ or, on average to $q_0-2\pi/L$ (while relaxing) which makes the system remain confined to the free energy range dominated by the quadratic terms of the Eq.2. At the time of relaxation, since it is a small change in the wave number that has to be accommodated, we consider it not to perturb the free energy profile in Fig.1 appreciably. Note that, the position of the free energy trough moves towards smaller A values besides the depth getting reduced as q approaches unity (i.e. q becomes larger). Thus, an allowed variation in q could make the system move to the quartic term dominated part of the free energy while amplitude gets reduced. But, the effective constancy of q does not allow for such opportunity. So, any form of amplitude modulations consistent to this excursion be reasonably found out from the linear part of the dynamics (Eq.1).
\par
The linearized system that we would consider from now on, as our model, for looking at amplitude modulations is
\bea
\psi_t+i\mu\psi=\psi_{xx} + (1-A_0^2)\psi, 
\eea
which might be complemented by the same expression of current density as in Eq.1, but, we do not require any explicit mention of this other equation in what follows. Note that, one can not simply drop nonlinear term from Eq.1 to get the linearized from of the dynamics. Instead, one has to replace the $|\psi|^2$ by the constant $A_0^2$ which is the amplitude of the pure SC phase at the free energy minimum. This form of the model preserves all the steady state properties of Eq.1. Recall the form of the pure SC solution of Eq.1 where the super-current density (measurable quantity) can give the wave number of the SC state as $j=q(1-q^2)$ and that also fixes the value of the constant $A_0^2$ in Eq.3 as a function of the current density. To find out the amplitude modulations from the proposed model we would look at the dynamics of the amplitude at a larger scale. Practically, amplitude variations happen over a much larger scale (proportional to ($\xi(T)$) \cite{kop}) and the intended results would show up at these larger scales. 
\par
Let us expand the order parameter $\psi$ as $\psi=A(X,\tau)[\psi^0+\epsilon\psi^1]$ where the amplitude of the order parameter $A(X,\tau)$ is a function of slow space and time scales $X$ and $\tau$ respectively. Take the N phase order parameter $\mu$ to be $O(\epsilon)$ and that is because we consider the associated $\mu$ profile to an amplitude modulation would be at the same scales. The small ($<1$) parameter $\epsilon$ is the ratio of the small to the large length scales i.e. the scales at which the phase of the stationary SC state and its amplitude vary respectively.  Since, the dynamics of the system in this resistive state is slow, the only time scale under effective consideration here is the slow one $t = \epsilon \tau$ and we introduce multiple length scales as $x=x+\epsilon X$. The standard tools of singular perturbations can now be brought to bear. Expanding up to $O(\epsilon)$ we get
\ber\nonumber
\epsilon\frac{\partial A(X,\tau)}{\partial \tau}[\psi^0+\epsilon\psi^1]&+&\epsilon A(X,\tau)\frac{\partial\psi^0}{\partial \tau}+\epsilon (i\mu\psi^0)\\\nonumber & =& A(X,\tau)\left[ \frac{\partial^2\psi^0}{\partial x^2} + \epsilon\frac{\partial^2\psi^1}{\partial x^2} \right] + \epsilon 2\frac{\partial A(X,\tau)}{\partial X}\left[ \frac{\partial\psi^0}{\partial x} + \epsilon\frac{\partial\psi^1}{\partial x} \right]\\ &+& \epsilon 2A(X,\tau)\frac{\partial^2\psi^0}{\partial X\partial x} + (1-A^2)A(X,\tau)[\psi^0+\epsilon\psi^1].
\eer
\par
Thus, at $O(1)$,
\bea
\psi_{xx}^0 +(1-A^2)\psi^0=0,
\eea 
is a zero eigenvalue equation and the solution is of the form $\psi^0=A(X,\tau)e^{iq(X,\tau)x}$. In the next order ($O(\epsilon)$),
\ber
\nonumber
&&\psi_{xx}^1 + (1-A^2)\psi^1=( \frac{\partial A(X,\tau)}{\partial \tau} + i\mu A(X,\tau)\\\nonumber &-& i2q(X,\tau)\frac{\partial A(X,\tau)}{\partial X} +ixA(X,\tau)\frac{\partial q(X,\tau)}{\partial \tau}\\ &+& (2xq(X,\tau)-2i)A(X,\tau)\frac{\partial q(X,\tau)}{\partial X})e^{iq(X,\tau)x}
\eer
The solvability, at this order, requires the coefficient of the secular term on the right hand side of the Eq.6 vanish. Considering the evolution of the wave vector to be independent of the amplitude dynamics we separate the equations for the amplitude and the wave vector as
\ber
\frac{\partial A}{\partial t} +i\epsilon\mu A - i2q\frac{\partial A}{\partial x}&=& 0\\
ix\frac{\partial q}{\partial t} +2(xq-i)\frac{\partial q}{\partial x}&=& 0.
\eer
In Eq.7 and 8 the smaller scales have been restored which effectively introduces an $\epsilon$ factor to the middle term of the Eq.7.
\par
The fixed points of the Eq.8 are $q=q_0$ (constant) and $q=i/x$. Considering $q=q_0 (=1-A_0^2)$, writing $\mu=-j_N x$ one can get a solution \cite{ari} for the full time dependent form of the Eq.7 as
\bea
A(x,t)=e^{-mx^2}[e^{\lambda^\prime x}e^{i\omega t}+e^{-\lambda^\prime x}e^{-i\omega t}]  
\eea
The constants in the above dynamic solution $m=\epsilon j_N/4q_0$, $\omega = \pm 2q_0\lambda^\prime$ and $\lambda^\prime$ is a real constant. This solution explicitly represents a dynamic PSC of the system. The solution is real at the origin, where due to superposition, the oscillatory part becomes $\cos{\omega t}$. Everywhere else, for nonzero $\lambda^\prime$, its a superposition of spirals (in time). A zero $\lambda^\prime$ is not allowed for dynamic PSCs because $\omega$ is proportional to $\lambda^\prime$ and no dynamic PSCs will form if A does not oscillate at the origin. Its important to note that, the oscillation frequency being proportional to wave number, the PSCs will be infrequent as the wave number approaches zero i.e. the system approaches the global minimum. This is not surprising, since, the height of the barrier it has to overcome is the same as the depth of the well $d=\frac{L\sigma}{2}(1-q_0^2)$ and, given that, no saddle is available to bypass the barrier in 1D. In a recent paper \cite{rub}, dynamic PSC generation has been shown as a consequence of the PT (parity and time reversal) invariance of the CGL which is also evident from the form of the above mentioned PSC solution. The solution is invariant under the joint symmetry operations of parity and complex conjugation reflecting the necessity of having PT invariance in the CGL for dynamic PSCs. The temporal oscillation of the solution irrespective of explicit consideration of the thermal fluctuations is interesting to note. Such localized oscillations were noticed numerically by Kramer et al \cite{kra} and latter on (also numerically) by others (see \cite{kop,wat,kra1} and references therein) within the resistive regime. These intrinsic dynamic PSCs are the dominant contributors to the phase loss at a region away from the $T_C$ where large thermal excitations are less probable \cite{kop}.
\par
To look at the static exact solutions captured by our model, consider the time independent part of Eq.7. In a simplified form this static part of the Eq.7
\bea
\frac{d}{dx}\ln{A(x)} = \frac{\epsilon }{2q_0}\mu (x),
\eea
relates the amplitude of the SC phase to the chemical potential and we argue that this is the central constraint that gives a selection for $\mu$ corresponding to a static modulation in A and vice versa. Following are the three solution pairs corresponding to some localized forms of the amplitude A
\begin{enumerate}
\item $A(x)=A_0 e^{-a_0x^2}, \mu=-\frac{4a_0q_0}{\epsilon\mu}x $
\item $ A(x)=A_0 {\text{sech}} {\alpha x}, \mu=-\frac{2q_0\alpha}{\epsilon }\tanh{\alpha x}$
\item $ A(x)=A_0e^{-a_0x^2}\cosh{\alpha x}, \mu=\frac{2q_0}{\epsilon }\left( -2a_0x+\alpha \tanh{\alpha x}\right )$
\end{enumerate} 
In the first pair of conjugate forms of $A$ and $\mu$, a Gaussian localized bubble of the SC phase is present on a predominantly normal phase over the entire wire which carries a constant Ohmic current. These Gaussian bubbles of the SC phases are similar in form to the numerically found out critical nuclei by Watts-Tobin et al \cite{wat,kop}. Such bubbles of SC phase are found to appear when one moves down the current from $j_2$ towards $j_c$ and the pre-existing N phase loses stability to an SC phase. If the bubbles are bigger than a critical size (critical nucleus), they grow and push the N phase out towards the end of the wire, otherwise, these SC bubbles vanish and the N phase remains stable. In the second pair of solutions, the hyperbolic tangent profile of $\mu$ represents an intervening normal phase connecting two states characterized by different constant electrochemical potentials. Thus, a localized amplitude modulated SC phase is again shown to coexist with a normal phase that basically connects two reservoirs of charges at two different electrochemical potentials. In Fig.2, we have plotted the third pair of the above solutions on the same scales as in the Fig.1 in ref.\cite{kra} for a comparison. The figure in the ref.\cite{kra} has been obtained numerically by integrating a nonlinear model similar to Eq.1. However, to our knowledge, there is no analytical expression existing for such structures till date. The remarkable resemblance of the analytical plot to the numerically obtained structure supports our claim that Eq.10 is the constraint that the amplitude and the phase dynamics remain decoupled during such structure formations. The parameters that have been used here to plot the analytic solution are $A_0=0.6$, $2q_0/\epsilon =0.35$, $a_0 =0.3$ and $\alpha=0.7$.
\par
A static PSC can have the form of $A(x)=A_0\tanh{\vert x\vert}$, which corresponds to the form of solution proposed by Langer and Ambegaokar for a small current density $j$ \cite{lan,kop}. At the origin, where the PSC is sitting, corresponding $\mu(x) = \frac{2q_0}{\epsilon }[1/\tanh{x}-\tanh{x}]$ will have a singularity. The analytic form we have derived here for the $\mu$ compares quite well with the numerically obtained one in ref.\cite{kop} (figure 18 in ref.\cite{kop}). Since, in the resistive regime, the SC and the N phase coexist, there could be some fronts present that separate these two phases \cite{kop}. For an example, consider an SC phase of the form $A(x)=A_0 e^x/(1+e^x)$. The Eq.3 immediately produces corresponding solution for $\mu$ as $\mu(x)= \frac{2q_0}{\epsilon }(1-e^x/(1+e^x))$. Thus, an SC phase on the right hand side is separated by a corresponding N phase on the left and vice versa by these fronts. The central constraint (Eq.10) correctly captures the numerically obtained structures in exact mathematical forms.
\par
To understand the implications of the existence of above mentioned amplitude modulations on the stability of the SC phase in the resistive regime, consider perturbing the fixed points of Eq.7 and 8, which are the above mentioned stationary local structures, with infinitesimal perturbations $\delta A$ and $\delta q$. One can readily linearize Eq.7 and 8, in a general way, for all of the above mentioned local structures to get
\bea
\frac{\partial}{\partial t} \begin{pmatrix} \delta A \\ \delta q \end{pmatrix} = \begin{pmatrix} {-(i \epsilon \mu_0+\frac{2kq_0}{u})} & {i2\frac{\partial A}{\partial x}} \\
0 & {ik(\frac{2}{x}+i\frac{2q_0}{u})} \end{pmatrix} \begin{pmatrix} \delta A \\ \delta q \end{pmatrix},
\eea
where $k$ is the characteristic wave number of the perturbations. The growth rates of the perturbations are $\lambda_{1,2}=-(i\epsilon\mu_0+\frac{2kq_0}{u}),(i\frac{2k}{ux}-\frac{2kq_0}{u})$. Real part of the growth rate $\lambda_{r}=-\frac{2kq_0}{u}$ is positive for opposite signs of $k$ and $q_0$. What interesting about this linear stability analysis is that, it shows in the presence of these amplitude modulations the SC state is locally unstable to perturbations. The constant amplitude SC phase is known to be locally stable and globally unstable in the resistive regime. The allowed amplitude modulations facilitate transitions of the SC state to a lower free energy regime by making it locally unstable. 
\par
To conclude, we have proposed a new model to find out exact analytic forms of the amplitude modulations associated with PSC formation and others. The model has been obtained on the basis of a hypothesis that the wave number of the SC phase remains constant during such structure formations. The solutions of the present model when compared with the already existing numerical results and experimental facts about localized modulations in the resistive phase vindicates our assumption of constancy of the wave number and resulting decoupling of the amplitude and phase dynamics. Linearity of the present model is an important aspect and its this linearity that allows us to get exact analytic solutions for a number of modulated states. We also have eventually derived a relation between the SC and N order parameters in the resistive regime which, we argue, is the basic constraint capable of producing all the observed stationary conjugate forms of the competing orders. An important consequence of the present analysis is that, it provides an exact measure of the free energy barrier in terms of the measurable super current density that the system is supposed to overcome while PSC forms. A revisit of the existing thermodynamics of such systems in the light of this new finding is worth doing in view of the large departures of the previously theoretically obtained numbers \cite{lan,mac} from the experimental ones.

\newpage

\begin{figure}
 \begin{psfrags} 
 \psfrag{f}[cc][][2.0][-90]{F}
 \psfrag{a}[cc][][2.0][0]{A}
  \includegraphics[scale=1,angle=-90]{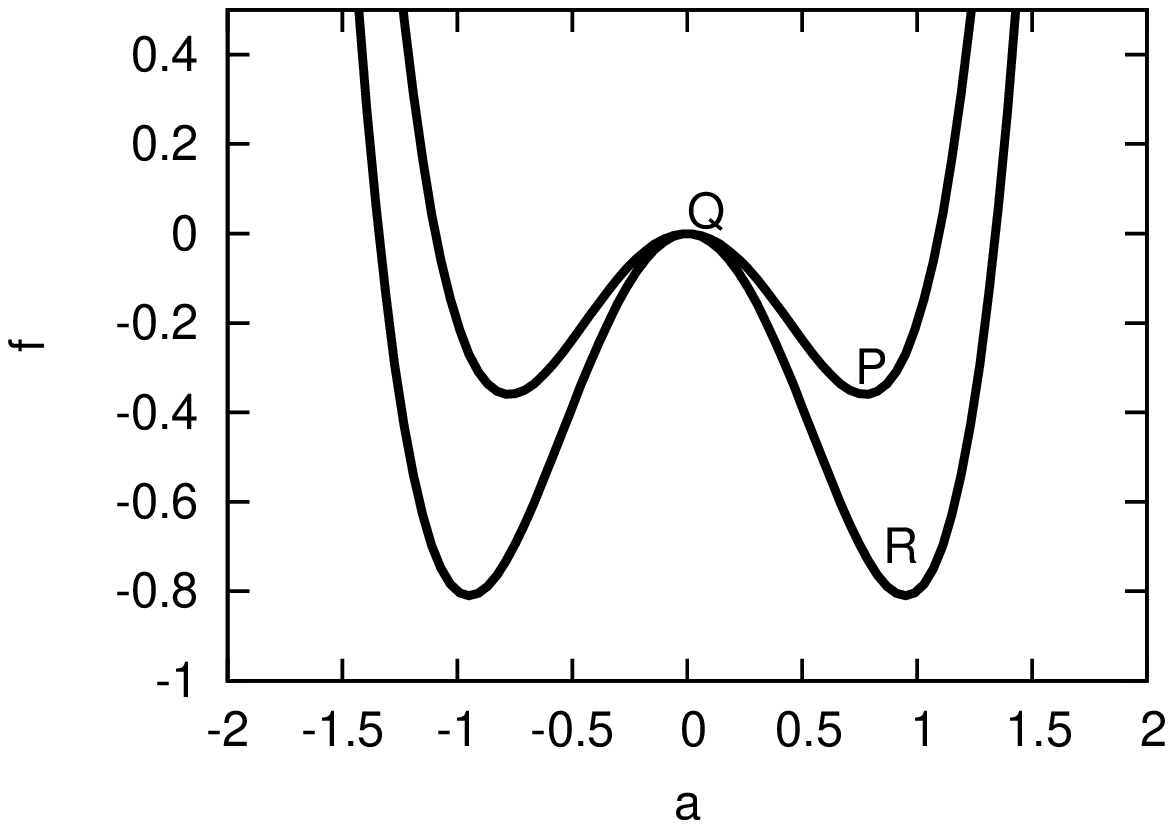}
 \end{psfrags}    
 \textsf{\caption[Figure 1]{Schematic diagram of the free energy F against amplityde A. During the formation of a PSC when the system loses a turn to lower the free energy the path the system would follow will be from a to o along the upper branch and them come down to b from o along the lower branch of the diagram.}}
\end{figure}

\begin{figure}
 \begin{psfrags} 
 \psfrag{b}[cc][][2.0][-90]{$\mu$}
 \psfrag{a}[cc][][2.0][-90]{A}
 \psfrag{x}[cc][][2.0][0]{X}
  \includegraphics[scale=1,angle=-90]{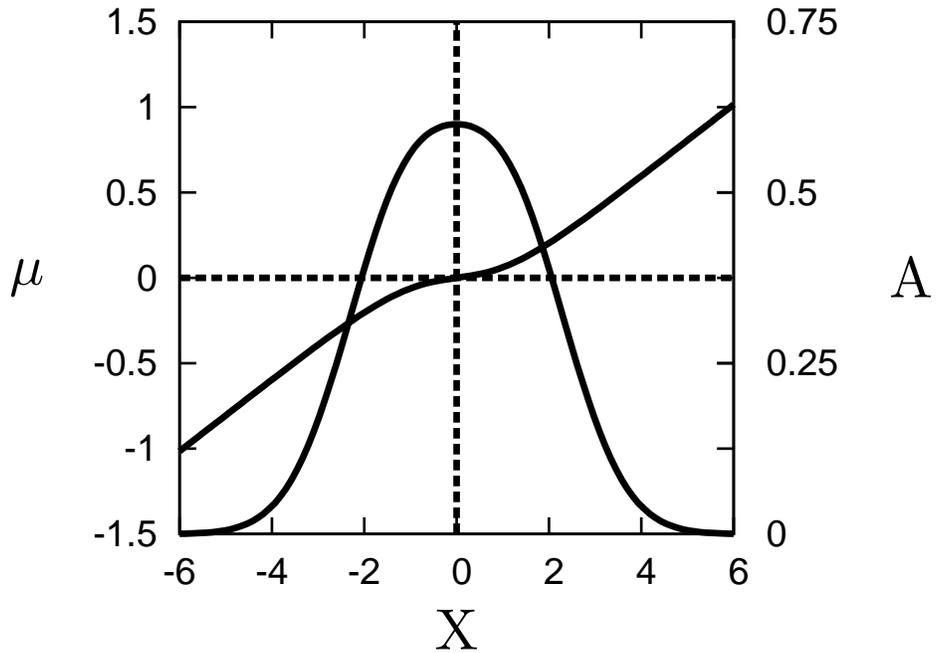}
 \end{psfrags}    
 \textsf{\caption[Figure 2]{A plot of the conjugate forms (relation 3 of the set showing static solutions of Eq.10) of localized amplitude and chemical potential variations for comparison with Fig.1 in ref.\cite{kra}.}}
\end{figure}

\end{document}